## On the flow of electrically charged particles in an elastic solid

**Jiashi Yang**
Department of Mechanical and Materials Engineering, University of Nebraska-Lincoln, Lincoln, NE 68588-0526, USA

**Abstract**
This paper is a specialization of a broad and complicated continuum theory [arXiv:2403.07582] to a relatively simple and useful case so that it is more reader friendly. A continuum theory of the flow of charged particles in an elastic solid is presented. It can describe the behavior of soft solid electrolytes and elastic semiconductors. It is nonlinear and is valid for large deformation and strong fields. The theory is derived from a three-continuum mixture model including a charged lattice continuum, a bound charge continuum for electric polarization, and an ideal fluid for the flow of mobile charges. The basic electromechanical laws are applied systematically to the model. The electric fields are quasistatic and are in SI units.

### 1. Introduction

When a deformable material is subjected to an electric field, the material experiences distributed electric body force and couple. The electric field also does work to the material during its polarization and conduction. Fundamental to the construction of a continuum theory for electroelastic materials is the derivation of the electric body force, couple and power due to the electric field. This can be achieved using the charged particle model in [1,2], or the multi-continuum model in [3-12] initiated by Tiersten, which may be viewed as a theory of mixtures [13]. The multi-continuum model has been used to construct a continuum theory of elastic solids carrying bound and mobile charges [7,12] which is rather complicated because of the inclusion of multi-physical couplings. In the following we specialize the theory in [7,12] to a simpler and yet still very useful situation by neglecting electromagnetic coupling and reducing the types of mobile charges from two to one. The electric field is treated as quasistatic and its coupling to the magnetic field is not considered. Thus the presentation below is much more readable and digestible than [7,12].

### 2. Multi-Continuum Model

The model consists of three constituents of the lattice continuum with charges fixed to it, the bound charge continuum which can displace a little from the lattice, and the mobile charge continuum which can flow through the lattice. The mobile charges may be ions, electrons, and holes [3-12]. They are modeled as an ideal fluid. Fields associated with the mobile charge fluid are indicated by a superscript $e$. The bound charge continuum is considered as massless.

Let the lattice be moving with $\mathbf{y} = \mathbf{y}(\mathbf{X},t)$ where $\mathbf{y}$ represents the present coordinates of the material particle of the lattice which was initially at $\mathbf{X}$ (material coordinates) in the reference state. The electric fields are $\mathbf{E}$, $\mathbf{D}$, and $\mathbf{P}$. The mechanical loads on the lattice are the body force **f** per unit mass and surface traction **t** per unit area. The lattice charge density is denoted by $\mu^l$ which experiences a force in an electric field. The external mechanical and electric forces on the lattice are shown in Fig 1.

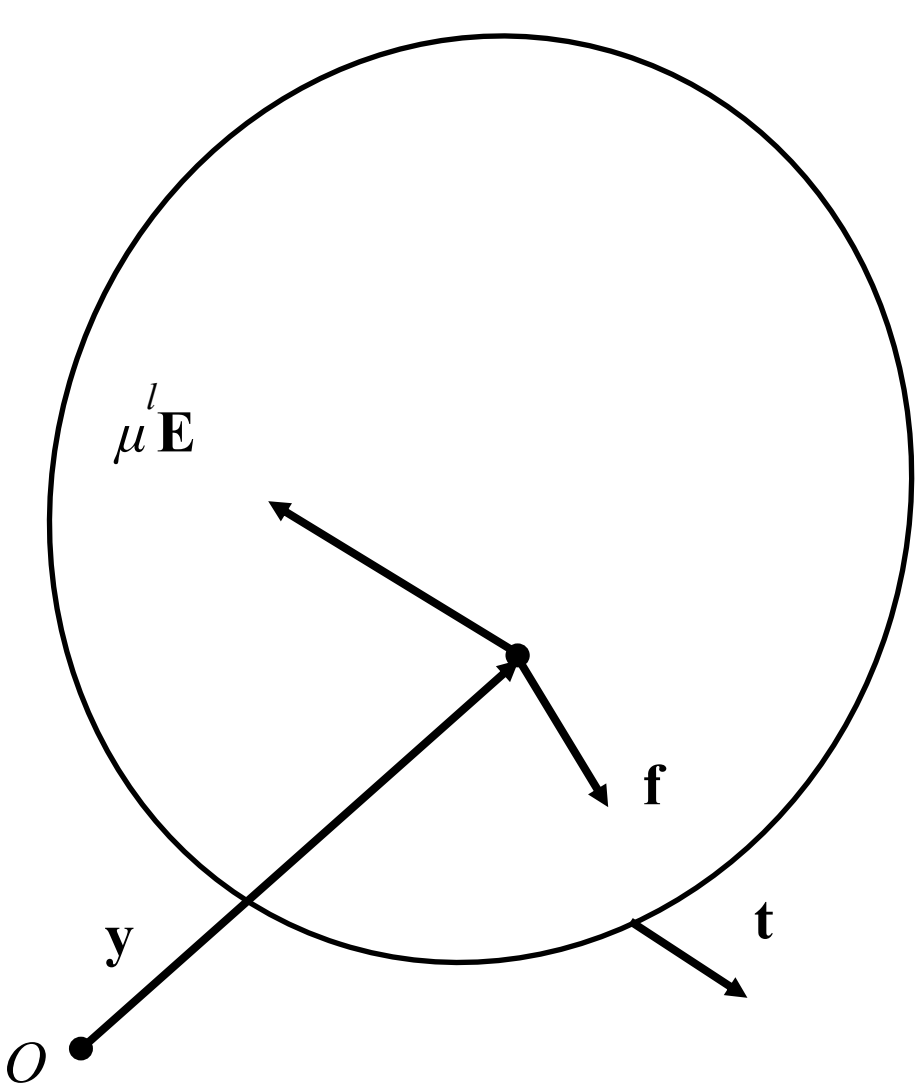

Fig. 1. External loads on the lattice. The interactions with the bound charge and the mobile charge are not shown.

The bound charge continuum can displace with respect to the lattice by a small displacement $\boldsymbol{\eta}(\mathbf{y},t)$ for describing electric polarization. The charge density of the bound charge continuum is $\mu^b$. We also denote

$$\mu^l(\boldsymbol{y}) + \mu^b(\boldsymbol{y}+\boldsymbol{\eta}) = \mu^r(\boldsymbol{y}), \tag{1}$$

which is referred to as the residual charge density. It is assumed that the small $\boldsymbol{\eta}$ satisfies [4]

$$\eta_{k,k} = 0. \tag{2}$$

With $\boldsymbol{\eta}$, the electric polarization $\mathbf{P}$ per unit volume is given by [4]

$$\boldsymbol{P} = \mu^b(\boldsymbol{y}+\boldsymbol{\eta})\boldsymbol{\eta} \cong \mu^b(\boldsymbol{y})\boldsymbol{\eta}, \tag{3}$$

$$\boldsymbol{\pi} = \boldsymbol{P}/\rho, \tag{4}$$

where $\boldsymbol{\pi}$ is the polarization per unit mass and only linear terms of the small $\boldsymbol{\eta}$ are kept. The external electric load on the bound charge continuum at $\mathbf{y}+\boldsymbol{\eta}$ is shown in Fig. 2.

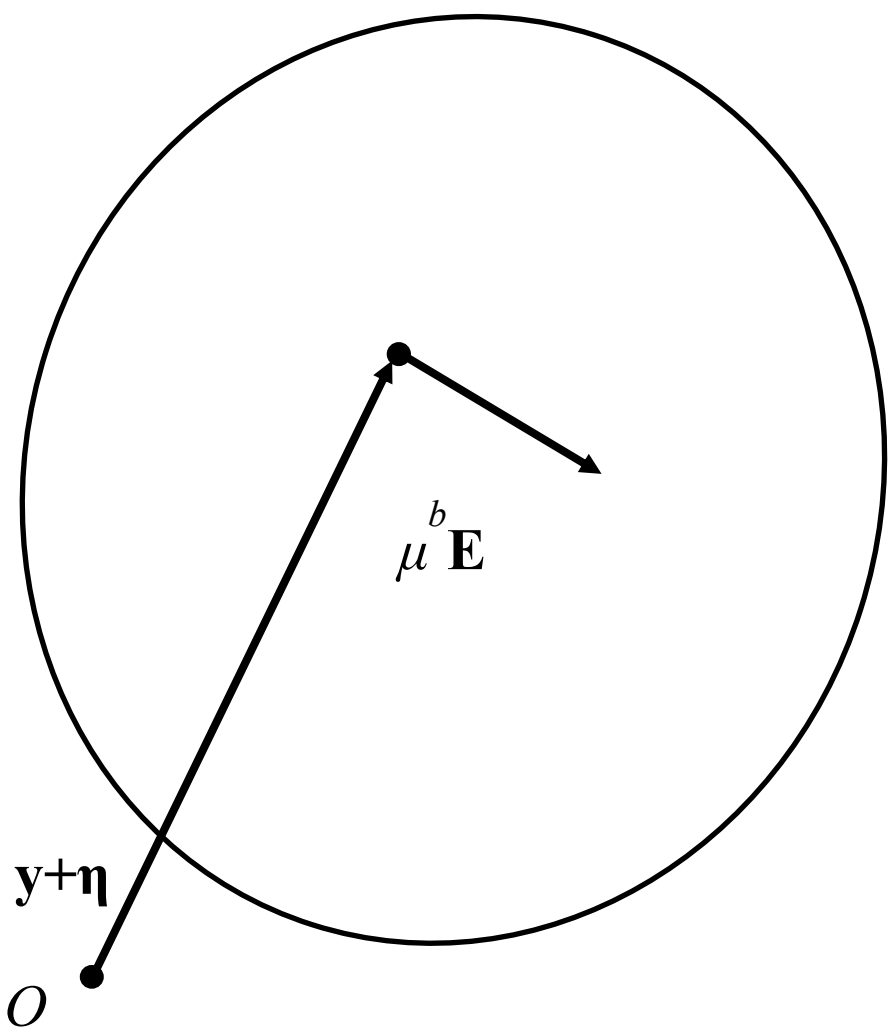

Fig. 2. External load on $\mu^b(\mathbf{y}+\boldsymbol{\eta})$.

The charge density of the mobile charge fluid is denoted by $\mu^e$. It interacts with the lattice through an effective local electric fields $\mathbf{E}^e$ [7]. The electric load per unit volume on the mobile charge fluid and its interaction with the lattice are shown in Fig. 3.

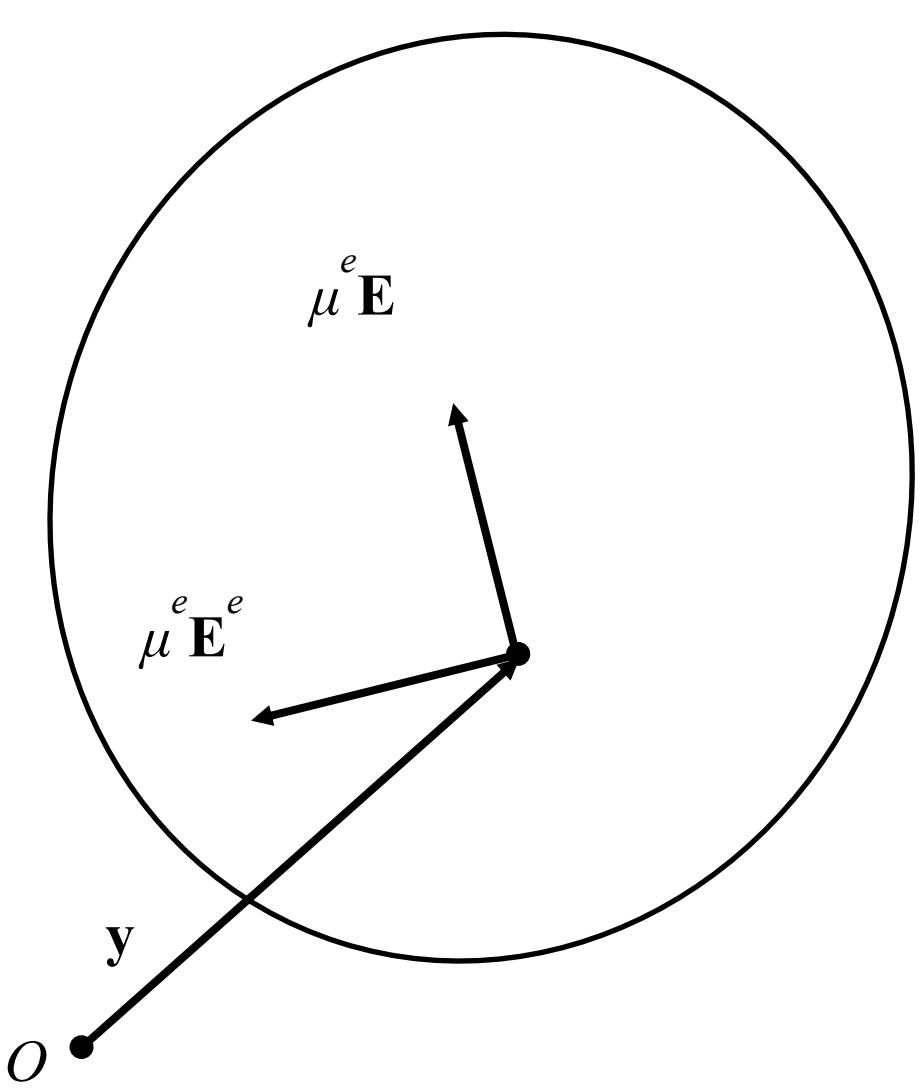

Fig. 3. Loads on the mobile charge fluid.

Let $\mathbf{v}$ and $\mathbf{v}^e$ be the velocity fields of the lattice and the mobile charge fluid. We denote the total charge density $\mu$ and current density $\mathbf{J}$ by [7,12]

$$\mu = \mu^r + \mu^e, \tag{5}$$

$$\boldsymbol{J} = \mu^r \boldsymbol{v} + \mu^e \boldsymbol{v}^e. \tag{6}$$

Then, from Figs. 1–3, through some algebra, it can be shown that the electric body force, couple and power on a unit volume of the lattice and mobile charge continua at **y** and the corresponding bound charge continuum at **y**+**η** can be written as

$$\boldsymbol{F}^E = \boldsymbol{P} \cdot \nabla \boldsymbol{E} + \mu \boldsymbol{E}. \tag{7}$$

The gradient operator in equation (7) is with respect to **y**. The electric body force can be written as

$$F_j^E = T_{ij,i}^E, \tag{8}$$

where $\mathbf{T}^E$ is the electric stress tensor

$$T_{ij}^E = P_i E_j + \varepsilon_0 E_i E_j - \frac{1}{2} \varepsilon_0 E_k E_k \delta_{ij}. \tag{9}$$

Similarly, the electric body couple on a unit volume of the combined continuum can be found to be [7,12]

$$\boldsymbol{C}^E = \boldsymbol{P} \times \boldsymbol{E}. \tag{10}$$

The electric body power on a unit volume of the combined continuum is given by [7,12]

$$\begin{aligned} W^E &= (\boldsymbol{P} \cdot \nabla \boldsymbol{E}) \cdot \boldsymbol{v} + \rho \boldsymbol{E} \cdot \dot{\boldsymbol{\pi}} + \boldsymbol{E} \cdot \boldsymbol{J} \\ &= \boldsymbol{F}^E \cdot \boldsymbol{v} + \rho \boldsymbol{E} \cdot \dot{\boldsymbol{\pi}} + \boldsymbol{J}' \cdot \boldsymbol{E}, \end{aligned} \tag{11}$$

where a superimposed dot is the material time derivative following the lattice, and the conduction current with respect to the moving lattice has the following expression:

$$\boldsymbol{J}' = \boldsymbol{J} - \mu \boldsymbol{v} = \mu^e (\boldsymbol{v}^e - \boldsymbol{v}). \tag{12}$$

**3. Integral Balance Laws**

Consider a spatial region $v$ with a boundary surface $s$ whose outward unit normal is **n**. $l$ is a closed curve. The integral forms of the relevant ones of Maxwell's equations for electrostatics are:

$$\oint_l \boldsymbol{E} \cdot \boldsymbol{dl} = 0, \tag{13}$$

$$\int_s \boldsymbol{n} \cdot \boldsymbol{D} ds = \mu. \tag{14}$$

The thermomechanical balance laws can be written for individual constituents or any combinations of them. The conservation of mass of the lattice (including the impurity and bound charge) is:

$$\frac{\partial}{\partial t} \int_v \rho \, dv = -\int_s \boldsymbol{n} \cdot \boldsymbol{v} \rho \, ds. \tag{15}$$

The conservation of charge for the mobile charge fluid is:

$$\frac{\partial}{\partial t} \int_v \mu^e \, dv = -\int_s \boldsymbol{n} \cdot \boldsymbol{v}^e \mu^e \, ds, \tag{16}$$

where, for simplicity, the exchange of charges among the constituents are not considered. We note that the conservation of mass for the mobile charge fluid is not independent to its conservation of charge because of the constant charge-to-mass ratio assumed.

The linear momentum equation for the lattice and bound charge continua may be written as

$$\begin{aligned} \frac{\partial}{\partial t} \int_v \rho \boldsymbol{v} dv &= -\int_s (\boldsymbol{n} \cdot \boldsymbol{v}) \rho \boldsymbol{v} dv + \int_s \boldsymbol{t} ds \\ &+ \int_v (\boldsymbol{P} \cdot \nabla \boldsymbol{E}) \, dv + \int_v [\rho \boldsymbol{f} + \mu^r \boldsymbol{E} - \mu^e \boldsymbol{E}^e] \, dv. \end{aligned} \tag{17}$$

The mobile charge fluid is assumed to be an ideal fluid with a pressure field $p^e$ [7,12]. Its linear momentum equations assumes the following form:

$$\frac{\partial}{\partial t}\int_v \rho^e \boldsymbol{v}^e dv = -\int_s (\boldsymbol{n}\cdot\boldsymbol{v}^e)\rho^e\boldsymbol{v}^e dv$$
$$+\int_s -p^e\boldsymbol{n}\,ds + \int_v [\rho^e\boldsymbol{f}^e + \mu^e\boldsymbol{E} + \mu^e\boldsymbol{E}^e]\,dv, \tag{18}$$

where $\mathbf{f}^e$ is the mechanical body force on the fluid. The addition of (17) and (18) gives the following linear momentum equation for all constituents together:

$$\frac{\partial}{\partial t}\int_v (\rho\boldsymbol{v} + \rho^e\boldsymbol{v}^e)dv = -\int_s [(\boldsymbol{n}\cdot\boldsymbol{v})\rho\boldsymbol{v} + (\boldsymbol{n}\cdot\boldsymbol{v}^e)\rho^e\boldsymbol{v}^e]dv$$
$$+\int_s (\boldsymbol{t} - p^e\boldsymbol{n})ds + \int_v [\rho\boldsymbol{f} + \rho^e\boldsymbol{f}^e + \boldsymbol{F}^E]\,dv. \tag{19}$$

The angular momentum equation of the combined continuum with all constituents is

$$\frac{\partial}{\partial t}\int_v \boldsymbol{y}\times(\rho\boldsymbol{v} + \rho^e\boldsymbol{v}^e)dv = -\int_s [(\boldsymbol{n}\cdot\boldsymbol{v})\boldsymbol{y}\times\rho\boldsymbol{v} + (\boldsymbol{n}\cdot\boldsymbol{v}^e)\boldsymbol{y}\times\rho^e\boldsymbol{v}^e]ds$$
$$+\int_s \boldsymbol{y}\times(\boldsymbol{t} - p^e\boldsymbol{n})ds + \int_v [\boldsymbol{y}\times(\rho\boldsymbol{f} + \rho^e\boldsymbol{f}^e + \boldsymbol{F}^E) + \boldsymbol{C}^E]\,dv. \tag{20}$$

The energy equation of the combined continuum of all constituents can be written as

$$\frac{\partial}{\partial t}\int_v \left(\frac{1}{2}\rho\boldsymbol{v}\cdot\boldsymbol{v} + \frac{1}{2}\rho^e\boldsymbol{v}^e\cdot\boldsymbol{v}^e\right)dv + \frac{\partial}{\partial t}\int_v (\rho\varepsilon + \rho^e\varepsilon^e)dv$$
$$= \int_s (\boldsymbol{t}\cdot\boldsymbol{v} - p^e\boldsymbol{n}\cdot\boldsymbol{v}^e - \boldsymbol{n}\cdot\boldsymbol{q})ds - \int_s \boldsymbol{n}\cdot\left(\boldsymbol{v}\frac{1}{2}\rho\boldsymbol{v}\cdot\boldsymbol{v} + \boldsymbol{v}^e\frac{1}{2}\rho^e\boldsymbol{v}^e\cdot\boldsymbol{v}^e\right)ds$$
$$-\int_s \boldsymbol{n}\cdot(\boldsymbol{v}\rho\varepsilon + \boldsymbol{v}^e\rho^e\varepsilon^e)ds + \int_v (\rho\boldsymbol{f}\cdot\boldsymbol{v} + \rho^e\boldsymbol{f}^e\cdot\boldsymbol{v}^e + \rho r)dv + \int_v W^E dv, \tag{21}$$

where $\varepsilon$ and $\varepsilon^e$ are the internal energy densities of the lattice continuum and the mobile charge fluid, $\mathbf{q}$ the heat flux vector, and $r$ the body heat source. Let the corresponding entropy densities be $\eta$ and $\eta^e$. Then the entropy inequality of all constituents together is

$$\frac{\partial}{\partial t}\int_v (\rho\eta + \rho^e\eta^e)dv$$
$$\geq -\int_s (\rho\eta\boldsymbol{v}\cdot\boldsymbol{n} + \rho^e\eta^e\boldsymbol{v}^e\cdot\boldsymbol{n})ds + \int_v \frac{\rho r}{\theta}dv - \int_s \frac{\boldsymbol{q}\cdot\boldsymbol{n}}{\theta}ds. \tag{22}$$

## 4. Differential Balance Laws

Using the divergence theorem and Stokes' theorem, introducing the Cauchy stress tensor $\boldsymbol{\tau}$ through $\mathbf{t} = \mathbf{n}\cdot\boldsymbol{\tau}$, we obtain the differential forms of the balance laws in equations (13)–(18) as

$$\nabla\times\boldsymbol{E} = 0, \tag{23}$$
$$\nabla\cdot\boldsymbol{D} = \mu, \tag{24}$$
$$\frac{\partial\rho}{\partial t} + \nabla\cdot(\rho\boldsymbol{v}) = 0, \tag{25}$$
$$\frac{\partial\mu^e}{\partial t} + \nabla\cdot(\mu^e\boldsymbol{v}^e) = 0, \tag{26}$$
$$\nabla\cdot\boldsymbol{\tau} + \rho\boldsymbol{f} + \boldsymbol{P}\cdot\nabla\boldsymbol{E} + \mu^r\boldsymbol{E} - \mu^e\boldsymbol{E}^e = \rho\frac{d\boldsymbol{v}}{dt}, \tag{27}$$
$$-\nabla p^e + \rho^e\boldsymbol{f}^e + \mu^e(\boldsymbol{E} + \boldsymbol{E}^e) = \rho^e\frac{d^e\boldsymbol{v}^e}{dt}, \tag{28}$$

where $d/dt$ and $d^e/dt$ are material time derivatives following the lattice and the mobile charge fluid, respectively. Equation (28) may be viewed as a generalized Euler's equation in fluid mechanics. Adding equations (27) and (28), we obtain

$$\nabla\cdot\boldsymbol{\tau} - \nabla p^e + \rho\boldsymbol{f} + \rho^e\boldsymbol{f}^e + \boldsymbol{F}^E = \rho\frac{d\boldsymbol{v}}{dt} + \rho^e\frac{d^e\boldsymbol{v}^e}{dt}, \tag{29}$$

which corresponds to equation (19). The differential forms of the angular momentum equation, energy equation and entropy inequality corresponding to equations (20)–(22) are found to be

$$\varepsilon_{kij}\tau_{ij} + C_k^E = 0, \tag{30}$$

$$\rho\frac{d\varepsilon}{dt} + \rho^e\frac{d^e\varepsilon^e}{dt} - \frac{p^e}{\rho^e}\frac{d^e\rho^e}{dt} = \tau_{ij}v_{j,i} + \rho E_i\frac{d\pi_i}{dt}$$
$$-\mu^e\boldsymbol{E}^e\cdot(\boldsymbol{v}^e-\boldsymbol{v}) + \rho r - q_{i,i}, \tag{31}$$

$$\rho\frac{d\eta}{dt} + \rho^e\frac{d^e\eta^e}{dt} \geq \frac{\rho r}{\theta} - \left(\frac{q_i}{\theta}\right)_{,i}. \tag{32}$$

**5. Constitutive Relations**

Eliminating $r$ from equations (31) and (32), we obtain the Clausius–Duhem inequality as

$$\rho\left(\theta\frac{d\eta}{dt} - \frac{d\varepsilon}{dt}\right) + \rho^e\left(\theta\frac{d^e\eta^e}{dt} - \frac{d^e\varepsilon^e}{dt}\right) + \frac{p^e}{\rho^e}\frac{d^e\rho^e}{dt} + \tau_{ij}v_{j,i} + \rho E_i\frac{d\pi_i}{dt}$$
$$-\mu^e\boldsymbol{E}^e\cdot(\boldsymbol{v}^e-\boldsymbol{v}) - \frac{q_i\theta_{,i}}{\theta} \geq 0. \tag{33}$$

The following free energy densities for different constituents can be introduced through the corresponding Legendre transforms:

$$F = \varepsilon - \boldsymbol{E}\cdot\boldsymbol{\pi} - \eta\theta, \quad F^e = \varepsilon^e - \theta\eta^e. \tag{34}$$

Then the energy equation in equation (31) and the Clausius–Duhem inequality in equation (33) take the following forms:

$$\rho\left(\frac{dF}{dt} + \theta\frac{d\eta}{dt} + \eta\frac{d\theta}{dt}\right) + \rho^e\left(\frac{d^eF^e}{dt} + \theta\frac{d^e\eta^e}{dt} + \eta^e\frac{d^e\theta}{dt}\right) - \frac{p^e}{\rho^e}\frac{d^e\rho^e}{dt}$$
$$= \tau_{ij}v_{j,i} - P_i\frac{dE_i}{dt} - \mu^e\boldsymbol{E}^e\cdot(\boldsymbol{v}^e-\boldsymbol{v}) + \rho r - q_{i,i}, \tag{35}$$

$$-\rho\left(\frac{dF}{dt} + \eta\frac{d\theta}{dt}\right) - \rho^e\left(\frac{d^eF^e}{dt} + \eta^e\frac{d^e\theta}{dt}\right) + \frac{p^e}{\rho^e}\frac{d^e\rho^e}{dt} + \tau_{ij}v_{j,i} - P_i\frac{dE_i}{dt}$$
$$-\mu^e\boldsymbol{E}^e\cdot(\boldsymbol{v}^e-\boldsymbol{v}) - \frac{q_i\theta_{,i}}{\theta} \geq 0. \tag{36}$$

We break the stress and polarization into recoverable and dissipative parts as

$$\boldsymbol{\tau} = \boldsymbol{\tau}^R + \boldsymbol{\tau}^D, \quad \boldsymbol{P} = \boldsymbol{P}^R + \boldsymbol{P}^D. \tag{37}$$

The recoverable parts satisfy

$$\rho\left(\frac{dF}{dt} + \eta\frac{d\theta}{dt}\right) + \rho^e\left(\frac{d^eF^e}{dt} + \eta^e\frac{d^e\theta}{dt}\right) - \frac{p^e}{\rho^e}\frac{d^e\rho^e}{dt} = \tau_{ij}^R v_{j,i} - P_i^R\frac{dE_i}{dt}. \tag{38}$$

Then the energy equation and Clausius–Duhem inequality reduce to

$$\rho\theta\frac{d\eta}{dt} + \rho^e\theta\frac{d^e\eta^e}{dt} = \tau_{ij}^D v_{j,i} - P_j^D\frac{dE_j}{dt} - \mu^e\boldsymbol{E}^e\cdot(\boldsymbol{v}^e-\boldsymbol{v}) + \rho r - q_{i,i}, \tag{39}$$

$$\tau_{ij}^D v_{j,i} - P_j^D\frac{dE_j}{dt} - \mu^e\boldsymbol{E}^e\cdot(\boldsymbol{v}^e-\boldsymbol{v}) - \frac{q_i\theta_{,i}}{\theta} \geq 0. \tag{40}$$

Equation (39) is the dissipation or heat equation. Equation (40) imposes restrictions on the dissipative parts of the constitutive relations.

For the recoverable fields satisfying equation (38), consider

$$F = F(v_{j,i}, \boldsymbol{E}, \theta), \quad F^e = F^e[(\rho^e)^{-1}, \theta]. \tag{41}$$

Since

$$v_{j,i} = X_{M,i}\frac{d}{dt}(y_{j,M}), \tag{42}$$

we can write equation (38) as

$$\left[X_{M,i}\tau_{ij}^{R} - \rho\frac{\partial F}{\partial(y_{j,M})}\right]\frac{d}{dt}(y_{j,M}) - \left(P_i^R + \rho\frac{\partial F}{\partial E_i}\right)\frac{dE_i}{dt}$$
$$-\rho\left(\eta + \frac{\partial F}{\partial \theta}\right)\frac{d\theta}{dt} - \rho^e\left(\eta^e + \frac{\partial F^e}{\partial \theta}\right)\frac{d^e\theta}{dt} + \frac{1}{\rho^e}\left(p^e + \frac{\partial F^e}{\partial(\rho^e)^{-1}}\right)\frac{d^e\rho^e}{dt} = 0, \tag{43}$$

which implies the following recoverable constitutive relations:

$$\tau_{ij}^{R} = \rho y_{i,M}\frac{\partial F}{\partial(y_{j,M})}, \quad P_i^R = -\rho\frac{\partial F}{\partial E_i},$$
$$\eta = -\frac{\partial F}{\partial \theta}, \quad \eta^e = -\frac{\partial F^e}{\partial \theta}, \quad p^e = -\frac{\partial F^e}{\partial(\rho^e)^{-1}}. \tag{44}$$

For rotational invariance (objectivity), $F$ can be reduced to a function of the following inner products and $\theta$:

$$C_{KL} = y_{i,K}y_{i,L}, \quad W_L = y_{i,L}E_i \ . \tag{45}$$

We will use the finite strain tensor $E_{KL}$ instead of the deformation tensor $C_{KL}$. Therefore, we take

$$F = F(E_{KL}, W_K, \theta),$$
$$E_{KL} = (C_{KL} - \delta_{KL})/2 = (y_{i,K}y_{i,L} - \delta_{KL})/2 = E_{LK}. \tag{46}$$

Then the first two of the constitutive relations in equation (44) become

$$\tau_{ij}^{R} = \rho y_{i,M}\frac{\partial F}{\partial E_{ML}}y_{j,L} + \rho y_{i,M}\frac{\partial F}{\partial W_M}E_j \ , \ P_i^R = -\rho y_{i,L}\frac{\partial F}{\partial W_L}. \tag{47}$$

It can be verified that the above constitutive relations obtained from a rotationally invariant energy density $F$ satisfy the angular momentum equation in equation (30). The dissipative parts of the constitutive relations may assume the following forms:

$$\boldsymbol{\tau}^D = \boldsymbol{\tau}^D(y_{j,M}, \boldsymbol{E}, \mu^e, \boldsymbol{E}^e, \theta \ldots),$$
$$\boldsymbol{P}^D = \boldsymbol{P}^D(y_{j,M}, \boldsymbol{E}, \mu^e, \boldsymbol{E}^e, \theta \ldots). \tag{48}$$

Similarly, the heat flux (Fourier's law) may be taken as

$$\boldsymbol{q} = \boldsymbol{q}(y_{j,M}, \boldsymbol{E}, \mu^e, \boldsymbol{E}^e, \theta, \theta_{,k} \ldots). \tag{49}$$

For the interactions among the constituents, while constitutive relations may be given in terms of the interaction represented by $\mathbf{E}^e$, they are usually given for $\mathbf{v}^e$–$\mathbf{v}$ for conduction (Ohm's law, see (12)):

$$\boldsymbol{v}^e - \boldsymbol{v} = \boldsymbol{V}^e(y_{j,M}, \boldsymbol{E}, \mu^e, \boldsymbol{E}^e, \theta, \theta_{,k} \ldots), \tag{50}$$

The dissipative constitutive relations are restricted by the Clausius–Duhem inequality in equation (40) and need to satisfy the requirements of rotational invariance too.

In addition, we have the following relationships:

$$\boldsymbol{D} = \varepsilon_0\boldsymbol{E} + \boldsymbol{P}. \tag{51}$$

An electric potential $\varphi$ may be introduced which satisfies (23).

## 6. Conclusions

In summary, the basic unknown fields are $\varphi$, $\rho$, $\rho^e$, $\mathbf{v}$, $\mathbf{v}^e$ and $\theta$. The basic field equations are (24)-(28) and (39). On the boundary surface of a finite body, the potential or the normal component of $\mathbf{D}$, the traction or velocity of the lattice continuum, and the pressure or normal component of the fluid velocity, and the temperature or the normal heat flux may be prescribed. When some of the couplings are neglected and the equations are linearized for small signals, the above equations are capable of producing the same dispersion curves of the linear and coupled plasma-acoustic waves in the literature [12].